*Research Article*

# Attitudinal Loyalty Manifestation in Banking: Cross-Buying Behavior and Customer Advocacy

[1]**Muhamad Bhayuta Yudhi Putera,** [2]**Melia Famiola**
[1,2]*School of Business and Management, Bandung Institute of Technology, Bandung, West Java, Indonesia.*

***Abstract:*** *This study in the banking industry examines the influence of attitudinal loyalty on customer advocacy and cross-buying behavior, alongside the moderating roles of Quality of Life and Corporate Social Responsibility (CSR) support in the CSR-fit and loyalty relationship. Employing Structural Equation Modeling (SEM) it reveals that higher attitudinal loyalty significantly boosts customer advocacy and propensity for cross-buying. The findings highlight the importance of nurturing customer loyalty through valuable and relevant offerings. Banks are advised to target customers with a high Quality of Life and engage with those who support CSR initiatives aligning with the bank's objectives, to enhance loyalty and deepen customer relationships.*

## I. INTRODUCTION

In the ever-evolving landscape of the Indonesian banking industry, traditional financial institutions face the dual challenge of adapting to digital transformation while maintaining a strong connection with their customer base. This dynamic environment necessitates a deep understanding of how various factors influence customer behavior, particularly in the context of Corporate Social Responsibility (CSR) practices. This research delves into the intricate interplay between a bank's CSR activities and customer behavior, focusing on cross-buying and customer advocacy as key indicators of customer loyalty. In this context, Corporate Social Responsibility (CSR) can serve as a marketing tool (Chahal, 1999; Sanclemente-Téllez, 2017) that enhances attitudinal loyalty among existing customers, enabling them to engage in cross-buying transactions. Attitudinal loyalty emerges as a result of CSR-fit, which refers to environmental or social activities that align with the company's values while also resonating with the hopes and empathy of its customers (Min et al., 2023; Ogunmokun et al., 2021; Ogunmokun and Timur, 2022).

The emergence of digital banking has revolutionized how customers interact with financial services, prompting traditional banks to innovate and adapt. In this context, CSR practices have become an essential tool for enhancing customer relationships and fostering loyalty. CSR's impact on the banking sector extends beyond mere compliance and philanthropy; it plays a crucial role in aligning a bank's operations with the evolving expectations of its stakeholders. The findings reveal that Corporate Social Responsibility (CSR) practices, particularly those focused on improving a company's image, can have a positive impact on shareholder value creation (Gallego-Álvarez et al., 2010). This is because investors can perceive the level of a company's dedication to sustainable development. These results highlight the potential of CSR as a marketing tool for companies (Brønn and Vrioni, 2001; Gallego-Álvarez et al., 2010; Sanclemente-Téllez, 2017). By engaging in CSR activities that enhance their image, companies can attract and retain investors who prioritize sustainable practices. This can lead to increased shareholder value and financial performance.

One strategy that has gained prominence in the banking sector is cross-buying, where banks offer existing customers additional products and services. This strategy has proven effective in not only boosting the bank's revenue but also in strengthening customer relationships. In the current research, we explore how CSR activities influence cross-buying behavior by enhancing attitudinal loyalty and a customer's emotional and psychological connection with the bank. This statement has also been supported and explored by other research studies (Min et al., 2023; Ogunmokun and Timur, 2022). Furthermore, cross-buying helps banks enhance customer retention and satisfaction (Bennett and Rundle-Thiele, 2002). This approach also allows banks to deepen their understanding of customers' financial preferences and behavior, enabling personalized and targeted offerings. Additionally, cross-buying enables banks to optimize operational efficiency and cost-effectiveness (Ogunmokun et al., 2021; Ogunmokun and Timur, 2022).

Another pivotal aspect of customer behavior in banking is customer advocacy, where loyal customers actively promote the bank through word-of-mouth and referrals. This study examines the relationship between attitudinal loyalty and customer advocacy, particularly in the context of a bank's CSR practices. We investigate how customers who perceive a bank's CSR activities as aligning with their values are more likely to advocate for the bank. Linking customer advocacy to the success of cross-buying within the banking sector highlights a symbiotic relationship that fuels both customer loyalty and diversified product engagement (Ogunmokun and Timur, 2022). Customer advocacy, characterized by enthusiastic customers promoting a brand, plays a pivotal role in fostering cross-buying, which involves customers purchasing multiple products or services from the same institution.

Moreover, the research considers the moderating effects of CSR support and quality of life on the relationship between CSR practices and attitudinal loyalty. It explores how customer support for CSR initiatives and their overall quality of life can influence their loyalty to a bank. This aspect is vital, as it reflects the broader impact of a bank's CSR activities on its customer base. Financial literacy also plays a crucial role in this dynamic. The study aims to integrate the variable of financial literacy into the research model, examining its influence on the relationship between CSR practices, attitudinal loyalty, and customer behaviors such as cross-buying and advocacy. Moreover, financial literacy enhances individuals' confidence in engaging with banking institutions (Hasan et al., 2021; Hsiao and Tsai, 2018). It enables them to navigate complex financial information, understand terms and conditions, and effectively compare offerings from different banks.

Bank XYZ's substantial 340-billion-rupiah CSR investment in 2023, significantly exceeding that of Indonesia's largest bank, underscores its strong commitment to CSR. This raises questions about the efficiency and potential overinvestment in its CSR initiatives. This research is pivotal for both the academic and banking sectors. For academia, it offers a comprehensive analysis of the factors that drive customer loyalty in banking, especially in the context of CSR. For practitioners, it provides insights into effective strategies for enhancing customer engagement and loyalty through CSR initiatives. This research addresses a critical problem in the banking industry: understanding the mechanisms through which CSR practices influence customer loyalty, particularly in the context of cross-buying and customer advocacy. It seeks to unravel the complexities of customer behavior in the banking sector, providing valuable insights for both academic research and practical application in the banking industry.

## II. LITERATURE REVIEW

### A) Bank XYZ

Bank XYZ, a notable entity in the Indonesian banking sector, has established itself as a key contributor to the country's economic growth, particularly through its support of micro, small, and medium-sized enterprises (MSMEs). Its extensive geographical reach and commitment to leveraging technology underscore its role as a modern, progressive financial institution. The bank's history and evolution reflect a deep-rooted commitment to serving its clients, particularly MSMEs, and its strategic focus on financial inclusion and technological advancements has positioned it as a significant player in Indonesia's financial landscape.

In the realm of Corporate Social Responsibility (CSR), Bank XYZ has developed a comprehensive program aligned with its Environmental, Social, and Governance (ESG) strategy. The bank's CSR activities, structured under various pillars, demonstrate its dedication to sustainable development and societal well-being. In the social sphere, XYZ has allocated substantial funds for initiatives aimed at enhancing community welfare and addressing basic human rights. These initiatives include educational support, disaster relief efforts, and medical equipment donations designed to provide immediate aid and foster long-term societal improvement.

Economically, XYZ has initiated empowerment programs for women's business groups and development programs for villages, focusing on both economic growth and social benefits. These programs support local economic development and community empowerment. Environmentally, XYZ's strategy encompasses sustainable resource management efforts, including waste management programs and initiatives to reduce greenhouse gas emissions. The adoption of green building practices further illustrates the bank's commitment to environmental sustainability.

XYZ's multifaceted CSR approach not only addresses current societal and environmental challenges but also contributes to sustainable development, reinforcing its role as a socially and environmentally responsible corporation. This commitment to CSR has garnered XYZ prestigious awards, including "The Best CSR of The Year" and "CEO CSR of The Year" at the Nusantara CSR Awards. These accolades are a testament to XYZ's impactful contribution to various social and environmental initiatives.

The bank's alignment with the Sustainable Development Goals (SDGs) is evident in the execution of over 2,000 programs through its CSR initiative. These programs target a range of goals, from poverty eradication and hunger reduction to gender equality and economic growth. XYZ's CSR initiatives not only provide immediate societal benefits but also focus on long-term sustainable development. The breadth of these activities and the recognition received highlight XYZ's role as a socially responsible entity, actively contributing to community well-being and environmental sustainability.

*B) CSR-fit Affects Attitudinal Loyalty*

This hypothesis suggests that the alignment between a bank's CSR initiatives and the expectations of customers (CSR-fit) can influence their propensity to engage in cross-buying behavior (Ogunmokun and Timur, 2022; Ogunmokun et al., 2021). CSR-fit refers to how well a company's CSR activities align with the values and interests of its customers (Becker-Olsen et al., 2006). When customers perceive a strong fit between a bank's CSR initiatives and their values, it enhances their positive attitudes toward the company. This positive attitude, in turn, leads to increased loyalty toward the bank and its offerings (Min et al., 2023).

Attitudinal loyalty then becomes a mediator in the relationship between CSR-fit and cross-buying behavior. Customers who have a higher level of attitudinal loyalty towards a bank are more likely to engage in cross-buying behavior, which involves purchasing products or services from multiple categories from the same bank (Bhatnagar et al., 2019; Larsson and Viitaoja, 2017).

**Hypothesis 1**: The higher the alignment between a Bank's CSR activities and its business, the greater the level of attitudinal loyalty.

*C) CSR-fit Affect Attitudinal Loyalty, Moderated by CSR Support and Quality of Life*

The second hypothesis aims to examine the link between a bank's CSR fit and attitudinal loyalty while also exploring how the level of CSR support from customers may influence the strength of this relationship. CSR-fit refers to the congruence between a bank's CSR activities or initiatives and the values, beliefs, and expectations of its stakeholders, particularly its customers. It reflects the extent to which the CSR efforts of a bank resonate with the values and interests of its target audience (Becker-Olsen et al., 2006). On the other hand, CSR support refers to the actual endorsement or approval that customers demonstrate towards a bank's CSR activities. It measures the degree to which customers actively support or appreciate the CSR initiatives undertaken by a company (Marin and Ruiz, 2007; Ramasamy et al., 2010).

When customers perceive a strong fit between a bank's CSR efforts and their values, it enhances their positive attitudes toward the bank. This alignment creates a sense of resonance and compatibility, leading to increased loyalty. However, the impact of CSR-fit on attitudinal loyalty is not the same for everyone. The quality of life that individuals experience plays a moderating role in this relationship (Asamoah et al., 2011; Ogunmokun et al., 2021). Quality of life refers to the overall well-being and satisfaction of an individual derived from various aspects of their lives, such as their physical and mental health, education, leisure time, and social belonging.

**Hypothesis 2:** Both customer support for the Bank's CSR activities and quality of life moderate the relationship between the alignment of the Bank's CSR activities with its business and attitudinal loyalty.

*D) Attitudinal Loyalty Affects Cross-Buying and Customer Advocacy*

When a bank's CSR initiatives align with customer values (CSR-fit), it can influence their cross-buying behavior. However, the impact of CSR-fit on cross-buying behavior is indirectly moderated by the level of CSR support demonstrated by customers. Higher levels of CSR support strengthen the association between CSR fit and cross-buying behavior, while lower levels of CSR support may weaken this relationship (Ogunmokun et al., 2021).

The relationship between attitudinal loyalty and customer advocacy is characterized by a positive correlation (Ogunmokun and Timur, 2022). This connection signifies that as customers develop a stronger emotional bond and allegiance to a particular brand, they are more inclined to engage in customer advocacy activities. Attitudinal loyalty goes beyond mere satisfaction, encompassing a deeper emotional connection, trust, and a positive attitude towards the brand Lawer and Knox (2006). When customers feel this strong affinity, they become brand advocates, willingly recommending the brand to others, sharing positive experiences, and actively supporting the brand's initiatives. This active promotion through word-of-mouth and positive engagement contributes to increased brand awareness, a positive reputation, and improved customer retention (Lawer and Knox, 2006).

Brand advocates, driven by their high attitudinal loyalty, actively participate in brand-related events, interact with the brand on various platforms, and promote the brand through positive word-of-mouth on social media and online forums (Lawer and Knox, 2006). Their willingness to recommend the brand to friends, family, and colleagues further strengthens the brand's reach and influence. In essence, fostering attitudinal loyalty among customers is crucial for cultivating a community of loyal advocates, which offers substantial benefits to the brand in terms of enhanced brand loyalty, positive brand perception, and organic growth through customer referrals. By nurturing this emotional connection, businesses can create a cycle of customer advocacy that reinforces brand loyalty and fuels long-term success.

**Hypothesis 3**: A higher level of attitudinal loyalty in customers significantly increases both the likelihood of engaging in cross-buying behavior and the propensity for customer advocacy.

*E) Attitudinal Loyalty Affect Customer Advocacy, Mediates by Cross-Buying*

The connection between attitudinal loyalty and customer advocacy is influenced by the process of cross-buying (Ogunmokun and Timur, 2022). In this context, cross-buying denotes customers' willingness to explore and purchase additional products or services from the same brand. As customers with strong attitudinal loyalty engage in cross-buying, they have the opportunity to experience the brand's diverse offerings and the quality of these extended products. This positive cross-buying experience reinforces their initial positive perception and emotional connection to the brand.

As cross-buying provides customers with positive experiences and exposure to more of the brand's offerings, it plays a pivotal role in amplifying customer advocacy. Satisfied customers, having experienced the brand's diverse product range, become even more enthusiastic advocates (Ogunmokun and Timur, 2022). Their heightened engagement in cross-buying drives them to share not only their initial purchase experiences but also the positive encounters they had with other products they discovered through cross-buying. As a result, cross-buying acts as a mediator between attitudinal loyalty and customer advocacy, strengthening the relationship and creating a cycle of customer advocacy. Through this process, the brand benefits from increased advocacy efforts, word-of-mouth promotion, and a growing base of loyal customers, all driven by their strong emotional attachment and attitudinal loyalty to the brand.

**Hypothesis 4**: Cross-buying mediates the relationship between attitudinal loyalty and customer advocacy.

## III. RESEARCH METHOD

*A) Method*

This study, classified into four groups based on objectives, benefits, time dimension, and data collection technique, falls under explanatory research. It aims to elucidate the deeper manifestation of customer advocacy as a function of attitudinal loyalty mediated by cross-buying behavior. It investigates the moderating effects of Quality of Life (QoL) and CSR support on the CSR-fit and loyalty relationship. Employing a quantitative approach, it formulates hypotheses grounded in existing theories and empirical studies and gathers primary data through questionnaires. The data is analyzed to predict relationships between variables, testing the hypotheses.

The study's significance lies in providing insights for both the banking industry and academia, particularly regarding CSR practices and their impact. By utilizing Structural Equation Modeling (SEM), a sophisticated statistical analysis method, the research examines complex relationships among variables. SEM integrates factor analysis and multiple regression analysis, enabling the evaluation of linkages between observed variables and latent constructs, thereby enriching the understanding of customer behaviors and attitudes in the banking sector. Partial least squares structural equation modeling (PLS-SEM) represents a synthesis of interdependence and dependence methodologies (Hair et al., 2018).

*B) Population and Sample*

This study focuses on investigating customer advocacy among Indonesian bank customers, emphasizing its importance in enhancing competitiveness and service improvement in the banking sector. Recognizing that customer advocacy transcends transactional interactions to encompass deeper engagement with a bank's array of services, the research aims to provide insights into this phenomenon (Ogunmokun and Timur, 2022; Lawer and Knox, 2006). The study employs a simple random sampling method to survey 191 bank customers, ensuring unbiased sampling for generalizable results. This sample size was selected for its statistical robustness, supporting reliable insights into customer advocacy trends in Indonesian banking.

Data is gathered through an online survey and structured questionnaire, enhanced by a preliminary pilot test for accuracy. Online data collection offers easy access, participant convenience, and a wider reach, aiming to increase response rates and reduce data gaps. Ethical considerations, including participant welfare, informed consent, and data confidentiality, are prioritized. The study's methodology aims to provide valuable insights and actionable recommendations for enhancing customer advocacy strategies in Indonesian banks. The pilot test's role is to refine the questionnaire, focusing on question clarity, survey flow, and time efficiency, thereby improving data capture efficiency.

*B) Questionnaire Design*

The questionnaire for this study is meticulously crafted to ensure ethical standards and comprehensive data collection. Participants are required to give informed consent, ensuring confidentiality and voluntary participation. The survey includes demographic queries such as age, gender, occupation, education level, and geographic location to analyze customer perceptions across diverse segments. It also assesses financial literacy through questions on financial concepts, exploring their impact on banking behaviors.

The study focuses on key variables: CSR-Fit, CSR support, Quality of Life, Attitudinal Loyalty, Customer Advocacy, and Cross-Buying behavior. These are integrated to delve into customer perceptions and behaviors in the Indonesian banking sector. The objective is to understand how these factors interact, offering strategic insights for enhancing customer experiences and strengthening the bank's competitive edge.

CSR-fit will serve as the independent variable, representing the alignment between a bank's corporate social responsibility initiatives and the values and beliefs of its customers (Becker-Olsen et al., 2006; Ogunmokun et al., 2021). CSR-fit is crucial in determining how well customers perceive the bank's efforts towards social and environmental causes and whether it positively influences their attitudes and behaviors towards the bank. In assessing CSR-Fit perception, a Likert scale ranging from 1 to 5 is employed to gauge various aspects. First, respondents are asked to rate the alignment of their bank's CSR activities with its brand image, services, or products (CFP1). A score of 1 indicates strong disagreement, while 5 signifies strong agreement. Second, participants are prompted to assess the degree of consistency between their bank's CSR initiatives and its brand, image, services, or products (CFP2). Lastly, respondents are encouraged to rate the extent to which their bank's CSR activities complement its brand image, services, or products (CFP3) using the same 1 to 5 Likert scale. These measurements collectively provide insight into how well a bank's CSR efforts align with its overall identity and offerings from the perspective of its customers.

CSR support and Quality of Life will be incorporated as moderator variables. CSR support refers to the level of awareness and support customers have for the bank's corporate social responsibility initiatives (Pérez and Rodríguez del Bosque, 2013; Ramasamy et al., 2010). In assessing CSR Support, various dimensions are considered, each measured on a Likert scale ranging from 1 to 5. First, respondents are asked to rate their belief that their bank should contribute to solving social issues (CSP1). A score of 1 indicates strong disagreement, while 5 signifies strong agreement. Second, participants are prompted to evaluate whether their bank allocates a portion of its budget for donations and social projects to improve conditions for less privileged groups in society (CSP2).

Furthermore, respondents are encouraged to rate the extent to which their bank provides monetary contributions for cultural and social events (CSP3) on the same scale. Additionally, participants are asked to assess whether their bank plays a role in supporting economically disadvantaged generations (CSP4). Lastly, they rate their perception of their bank's commitment to enhancing the overall well-being of the community (CSP5).

Quality of life refers to the comprehensive assessment of an individual's overall well-being and life satisfaction, considering various aspects of their physical, mental, social, and emotional experiences (Costanza et al., 2007; Liu and Zhou, 2009). As moderators, these variables can influence the strength or direction of the relationship between CSR-fit and Attitudinal Loyalty and Cross-buying behavior, shaping the impact of CSR-fit on these dependent variables. In measuring Quality of Life (QoL), several dimensions are taken into account, with each dimension rated on a Likert scale ranging from 1 to 5. Firstly, respondents are asked to rate the extent to which their life aligns with their dreams (QoL1). A score of 1 signifies strong disagreement, while 5 indicates strong agreement. Secondly, participants are prompted to evaluate the overall quality of their life (QoL2) and their satisfaction with it (QoL3). Then, respondents are encouraged to rate the degree to which they have achieved important life goals and desires (QoL4). Lastly, they are asked to consider whether, given the chance, they would make significant changes in their life if they could live it again (QoL5). These measurements collectively provide insights into how individuals perceive the quality and contentment in their lives, reflecting their aspirations and contentment levels.

Attitudinal Loyalty and Cross-buying behavior will be assessed as mediator variables. Attitudinal Loyalty represents customers' emotional attachment and commitment to the bank (Bandyopadhyay and Martell, 2007; Liu and Zhou, 2009). When evaluating Attitudinal Loyalty, three key dimensions are reflected in three indicators which are Cognitive, Affective, and Conative. Respondents are asked to rate each dimension on a Likert scale ranging from 1 to 5. Firstly, the cognitive dimension assesses customers' cognitive beliefs about their bank. Respondents are prompted to rate their overall belief in the excellence of the banking and financial services offered by their bank (ALO1), the overall quality of services (ALO2), the stress-free nature of using their bank (ALO3), and the bank's provision of a safe and conducive environment for financial transactions (ALO4). Secondly, the Affective dimension gauges emotional attachment and preferences. Participants are encouraged to rate their preference for their bank over comparable institutions (ALN1), the appeal of using their bank (ALN2), the importance of using their bank (ALN3), the enjoyment derived from using their bank (ALN4), and the pleasant experience provided by their bank (ALN5). Lastly, the conative dimension evaluates behavioral intentions and commitment. Respondents are asked to rate their intention to continue using their bank's services and products, even when better options are available elsewhere (ALE1), their intention to remain a customer of their current bank rather than switching to a new one (ALE2), and their intention to expand their usage of their bank's services (ALE3).

On the other hand, cross-buying behavior indicates their tendency to utilize multiple banking products and services from the same provider (Ogunmokun et al., 2021; Ramaseshan et al., 2017). Assessing cross-buying behavior involves considering four key dimensions, each rated on a Likert scale ranging from 1 to 5. Firstly, respondents evaluate their past use of more than one product or service from their bank, reflecting their historical engagement with the bank's offerings (CBB1). Secondly, it prompts respondents to assess their current utilization of multiple products or services from their bank, offering insights into their ongoing engagement (CBB2). Additionally, respondents reflect on whether their transaction volume with their bank has increased, indicating changes in their banking activity over time (CBB3). Lastly, respondents rate whether they choose their bank over others when purchasing additional banking services. Collectively, these dimensions provide a comprehensive view of customers' cross-buying behavior, encompassing their historical, current, and potential usage patterns within their banking relationship.

Customer Advocacy will serve as the dependent variable, reflecting customers' active promotion and recommendation of the bank's services to others. This variable captures the extent to which customers are satisfied, loyal, and willing to endorse the bank to their social circles. Customer Advocacy is influenced by the interplay of CSR-fit, Attitudinal Loyalty, and Cross-buying behavior, indicating the significance of these factors in driving positive customer behavior (Ogunmokun and Timur, 2022).

In the assessment of Customer Advocacy, four essential dimensions are considered, with each dimension being rated on a Likert scale ranging from 1 to 5. Firstly, respondents express their inclination to exclusively choose banking services from their bank, highlighting their loyalty and preference for their current bank over alternatives (CA1). Secondly, it captures respondents' active support and advocacy for their bank, reflecting their willingness to promote and champion the bank's offerings (CA2). Moreover, it assesses the extent to which respondents are willing to disseminate positive perceptions about their bank, signifying their role as brand advocates who share their positive experiences (CA3). Lastly, it delves into respondents' readiness to provide beneficial feedback for their bank, indicating their proactive engagement in helping the bank improve and excel (CA4).

Additionally, financial literacy will be considered as a control variable in the research. Financial literacy refers to customers' understanding of financial concepts and principles (Bannier and Schwarz, 2018). By including financial literacy as a control variable, the research seeks to account for its potential impact on the relationships between the main variables, ensuring that any observed effects are not confounded by differences in financial knowledge among participants.

We are using the "Big Three" financial literacy question to measure this variable (Lusardi and Mitchell, 2011). The questionnaire delves into specific financial scenarios to assess practical financial knowledge. One scenario presents a savings account with a fixed interest rate (Q1). Respondents are asked to predict the amount of their savings after five years, considering the interest rate. This question not only measures mathematical proficiency but also the ability to comprehend and apply interest rate concepts. Another scenario introduces the dynamics of inflation (Q2). Respondents are prompted to consider how inflation impacts the purchasing power of their savings. This question assesses whether individuals understand the erosion of the real value of money due to inflation, a fundamental aspect of financial literacy. The final question pertains to investing in stocks (Q3). It evaluates respondents' understanding of risk and return in investment decisions. Specifically, it inquires whether individuals believe that buying one share of a company's stock is inherently safer compared to diversifying investments across multiple stocks or mutual funds.

*C) Data Analysis*
A meticulous approach to data analysis is adopted, beginning with a pilot test for questionnaire validation, followed by data preparation for cleaning and organizing the raw data. Descriptive analysis is employed to summarize the data's primary characteristics. Partial Least Squares Structural Equation Modeling (PLS-SEM) is used to investigate the relationships between variables, offering insights into the study's hypotheses. We employ Structural Equation Modeling (SEM), a technique suited for evaluating complex relationships between variables, each comprising multiple indicators Hair et al. (2018). SEM's dual capacity to assess the measurement model (relations between observed indicators and latent variables) and the structural model (interconnections among latent constructs) allows for an in-depth examination of both direct and indirect effects. This approach acknowledges measurement error, enhancing the analysis's accuracy and robustness, and facilitates hypothesis testing involving mediation, moderation, and multiple variables.

This study employs Partial Least Squares Structural Equation Modeling (PLS-SEM) for its predictive capabilities and suitability for datasets with non-normal distributions, small sample sizes, and complex models. PLS-SEM is efficient in handling missing values and diverse measurement scales, offering robust solutions and predictive insights. Data analysis is facilitated by the SEMinR package in RStudio, which provides a user-friendly interface for SEM implementation. This package simplifies the modeling process, enhancing SEM's accessibility for researchers.

Overall, the research combines the analytical power of PLS-SEM with the user-friendly SEMinR package, ensuring a thorough and precise analysis. This methodological approach is instrumental in deriving insights into customer advocacy, satisfaction, and cross-buying behavior in the banking sector.

## IV. RESULTS AND DISCUSSION

Of the 240 initial survey responses from Bank XYZ customers, 50 were excluded due to participants not thoroughly reading the questionnaire, indicated by a zero-standard deviation in responses, suggesting no variability. The final data set thus comprised 191 valid responses, including 30 from bank employees, an aspect considered in the analysis. Collinearity diagnostics in the structural model showed all Variance Inflation Factor (VIF) values below the threshold of 5, confirming the management of collinearity concerns in the model.

Proceeding to the next step, the assessment focused on the significance and relevance of structural model relationships. This involved scrutinizing the confidence intervals to determine if they contained zero, thereby indicating insignificance. The evaluation revealed several noteworthy findings (see Table 1). First, the relationship from CSR-Fit to Attitudinal Loyalty was deemed non-significant due to the presence of zero within the confidence intervals. Similarly, the moderation effects of CSR Support and Quality of Life in the relationship between CSR-Fit and Attitudinal Loyalty were not deemed significant, as zero fell within their respective confidence intervals.

Conversely, the relationships from Attitudinal Loyalty to both Cross-Buying and Customer Advocacy were deemed significant, with no zero falling within their confidence intervals. However, the mediation effect of Cross-Buying in the relationship between Attitudinal Loyalty and Customer Advocacy was found to be non-significant, as zero was present within the confidence intervals. These results provide a comprehensive understanding of the significance and relevance of each structural relationship within the model.

In our study, we conducted a multi-group analysis to examine potential differences in the structural model between groups with varying levels of financial literacy. This analysis explored the impact of constructs like CSR-Fit, CSR Support, Quality of Life, and Attitudinal Loyalty on outcomes such as Cross-Buying and Customer Advocacy. However, the results indicated no significant differences between the groups of lower and higher financial literacy, implying that the hypothesized relationships within the structural model are consistent across different financial literacy levels. This finding highlights the model's robustness and general applicability, suggesting that the relationships hold true regardless of the participants' financial literacy.

The result presents Structural Equation Modeling (SEM) analyses, revealing the explanatory power ($R^2$) and adjusted explanatory power ($AdjR^2$) for key constructs within the attitudinal loyalty framework. Attitudinal Loyalty demonstrates a substantial $R^2$ of 0.391, indicating that approximately 39.1% of the variance in this construct is accounted for by the model. The corresponding $AdjR^2$ of 0.375 suggests that the model's explanatory strength remains robust after adjusting for the number of predictors.

Moving to the context of customer advocacy, the model highlights a notably high $R^2$ of 0.835, signifying an impressive 83.5% explanatory power. The robustness of this relationship is affirmed by the equally high $AdjR^2$ of 0.832. Interestingly, CSR-Fit exhibits a positive coefficient of 0.164, underscoring its role in influencing customer advocacy. However, CSR Support and the interaction term (CSR-Fit*CSR Support) with customer advocacy demonstrate less pronounced relationships, with respective coefficients of 0.327 and -0.025.

In the context of cross-buying behavior, the model elucidates an $R^2$ of 0.626, denoting that approximately 62.6% of the variance in cross-buying is explained by the specified constructs. The $AdjR^2$ of 0.624 reinforces the model's effectiveness. While Quality of Life and CSR-Fit*Quality of Life exhibit positive coefficients of 0.278 and 0.035, suggesting their influence on cross-buying, other relationships in this domain, such as CSR Support with cross-buying, appear less prominent. The interaction among these constructs offers a meaningful understanding of the complex relationship between attitudinal loyalty, customer advocacy, and cross-buying activities in the examined context.

### A) *Moderation Effect*

We tested two moderation effects in the relationship between CSR-fit and Attitudinal Loyalty. The examination of moderation effects in the structural model focused on two key relationships: CSR-Fit to Attitudinal Loyalty moderated by CSR Support and CSR-Fit to Attitudinal Loyalty moderated by Quality of Life. The results, as depicted in Table 1, explicitly indicate that both of these moderation effect relationships are not statistically significant. This conclusion is drawn from the observation that zero is encompassed within the confidence intervals for both relationships, signifying a lack of statistical significance. Our findings are further substantiated by the graphical representation in the following Figure 1 and Figure 2. The figure illustrates that, across the -1-standard deviation (SD) to +1 SD range for each moderating variable, there are no discernible differences in the impact on Attitudinal Loyalty. This visual representation reinforces the statistical insignificance observed in the moderation effect test results. In essence, these outcomes underscore the absence of a substantial moderating influence of CSR Support and Quality of Life on the relationship between CSR-Fit and Attitudinal Loyalty within the structural model. These results offer a nuanced comprehension of the constrained moderating impacts within the context of the scrutinized relationships, delivering valuable insights into the comprehensive evaluation of the structural model.

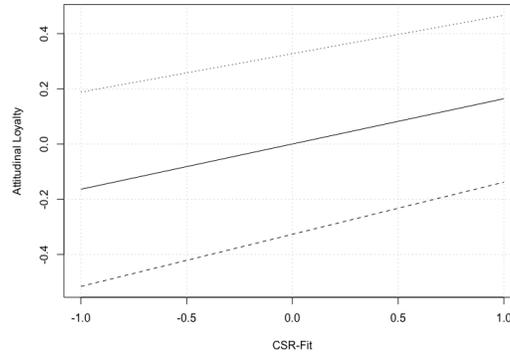

**Figure 1: Moderation Effect – CSR Support**

*B) Mediation Effect*

The examination of mediation effects in the current study encompassed two pivotal relationships: the mediation of Attitudinal Loyalty to Customer Advocacy by Cross-Buying Behavior and the mediation of CSR-Fit to Customer Advocacy through Attitudinal Loyalty. However, the outcomes, as indicated by the results presented in Table 1, reveal that both of these relationships lack statistical significance, as denoted by the presence of zero within the confidence intervals.

Interestingly, despite the insignificance of the direct effect of CSR-Fit on Attitudinal Loyalty, a noteworthy observation emerges—the direct relationship between Attitudinal Loyalty to Customer Advocacy attains statistical significance. These findings offer a comprehensive perspective on the nuanced interplay between these constructs, underscoring the importance of exploring various pathways within the structural model to discern their intricate associations. Hypothesis 1, exploring the relationship between CSR activities' alignment with business and attitudinal loyalty, was rejected, suggesting complexities in how CSR alignment influences customer loyalty. Hypotheses 2, examining the moderating roles of customer support for CSR and quality of life, also saw rejection, indicating these factors do not significantly alter the CSR-loyalty dynamic.

Conversely, Hypotheses 3, proposing positive links between attitudinal loyalty and both cross-buying and customer advocacy, was confirmed. These outcomes underline the crucial role of attitudinal loyalty in enhancing customer behaviors beneficial to the bank. However, Hypothesis 4, positing cross-buying as a mediator between attitudinal loyalty and customer advocacy, was not supported. This suggests a direct, significant impact of attitudinal loyalty on customer advocacy, independent of cross-buying.

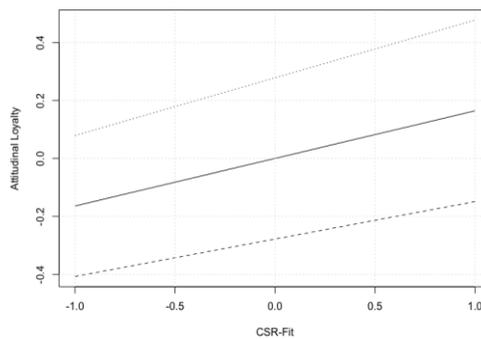

**Figure 2: Moderation Effect – Quality of Life**

### V. DISCUSSION

**The relationship between a Bank's CSR activities and its business does not significantly impact the level of attitudinal loyalty among its customers.**

The study indicates that the alignment of a bank's CSR activities with its business operations does not significantly influence customer attitudinal loyalty at Bank XYZ, contrary to our initial hypothesis. This suggests that CSR-fit may not be a crucial factor for Bank XYZ's customers in determining their loyalty.

In, subjects were categorized into banking and commercial banking, with banking customers assumed to exhibit greater CSR sensitivity. Although our research demographic closely aligns with commercial banking customers in Ogunmokun's study, our findings diverge, indicating similarities with customers characterized by heightened CSR sensitivity. Notably, (Ogunmokun and Timur, 2022) asserted that, in such circumstances, CSR-fit does not significantly impact attitudinal loyalty as customers with this persona perceive CSR-fit as a hygiene factor, a given expectation. Our study extends this perspective, revealing that, in the context of our research, customers evaluate CSR-fit as less pivotal than the quality of the product itself in determining loyalty. This does not imply that customers do not value a robust CSR-fit from Bank XYZ; rather, CSR-fit is not deemed the primary determinant of loyalty.

In various research contexts, brand image has been identified as a substantial precursor to customer behavior, encompassing attitudinal loyalty Lin et al. (2019). Moreover, distinct CSR strategies, such as adopting a more philanthropic approach rather than relying on sponsorship and cause-related marketing, coupled with efforts to diminish psychological distance, have been proposed to amplify the impact of CSR-fit on customer behavior (Lii et al., 2013). This body of literature underscores the multifaceted nature of factors influencing customer behavior, emphasizing the need for a comprehensive understanding of the interplay between CSR-fit, brand image, and diverse CSR strategies in shaping attitudinal loyalty.

**Table 1: Hypothesis Testing**

|  | Original Est, | Boot-strap Mean | Boot-strap SD | T Stat | 2,5% CI | 97,5% CI |
|---|---|---|---|---|---|---|
| CSR-Fit -> Attitudinal Loyalty | 0,164 | 0,161 | 0,088 | 1,860 | -0,007 | 0,337 |
| CSR-Fit -> Customer Advocacy | -0,070 | -0,067 | 0,041 | -1,699 | -0,146 | 0,014 |
| CSR Support -> Attitudinal Loyalty | 0,327 | 0,340 | 0,085 | 3,848 | 0,175 | 0,506 |
| CSR-Fit*CSR Support -> Attitudinal Loyalty | -0,025 | -0,023 | 0,054 | -0,460 | -0,136 | 0,082 |
| Quality of Life -> Attitudinal Loyalty | 0,278 | 0,294 | 0,076 | 3,641 | 0,146 | 0,442 |
| CSR-Fit*Quality of Life -> Attitudinal Loyalty | 0,035 | 0,039 | 0,068 | 0,511 | -0,081 | 0,191 |
| Attitudinal Loyalty -> Cross-Buying | 0,791 | 0,799 | 0,027 | 29,440 | 0,741 | 0,846 |
| Attitudinal loyalty -> Customer Advocacy | 0,898 | 0,900 | 0,054 | 16,595 | 0,791 | 1,004 |
| Cross-Buying -> Customer Advocacy | 0,059 | 0,058 | 0,055 | 1,069 | -0,047 | 0,167 |

Contrary to expectations, we found that CSR-Fit does not emerge as a predictor of customer advocacy in the investigated scenarios. The discernment of consumers regarding CSR domains is intricate, revealing a nuanced interplay influenced by personal values and opinions. Marketers are, therefore, advised to consider a corporation's core business industry context and align marketing strategies with the values and opinions of their target customers. Furthermore, the role of consumer skepticism emerges as a critical factor, indicating that individuals with heightened skepticism levels are less likely to respond positively to Customer Relationship Management (CRM) campaigns in comparison to their less skeptical counterparts (Skarmeas and Leonidou, 2013). Exploring an alternative mediation model, (Chung and Lee, 2022) found that CSR-fit positively influences behavior, specifically purchase intention, mediated through heightened perceived integrity and a positive attitude toward apology statements. Pérez and Rodríguez del Bosque's (2013) research highlights that psychological features are more effective than demographic factors in explaining customers' perceptions of CSR. Their finding underscores the importance of understanding the psychological drivers behind customer attitudes towards corporate social responsibility. These findings underscore the complexity of consumer responses to CSR initiatives, emphasizing the need for tailored and context-specific marketing strategies.

**Customer support for the Bank's CSR activities does not moderate the relationship between the alignment of the Bank's CSR activities with its business operations and the level of attitudinal loyalty.**

This study also does not uncover a significant moderating role for customer support in the relationship between the alignment of a Bank's CSR activities and its business, and attitudinal loyalty. These results contradict our hypothesis, positing that customer support moderates the association between the alignment of the Bank's CSR activities and its business and attitudinal loyalty. Despite this, the research model does indicate a positive relationship between CSR support and attitudinal loyalty.

The findings of this study diverge from Ogunmokun et al. (2021), who asserted that in their research, CSR support significantly serves as an underlying psychological attribute explaining the relationship between CSR fit and attitudinal loyalty in banking customers. However, in contrast, this study does not support the notion that CSR fit is a significant antecedent to attitudinal loyalty. The absence of CSR fit as a precise antecedent to attitudinal loyalty contributes to this situation. In a broader research context, Ramasamy et al. (2010) note that CSR support's significance is contingent on motivational factors influenced by societal values, highlighting the importance of a customized CSR strategy. Additionally, (Ginder et al., 2021) emphasize the need for internal-external congruence between CSR positioning and customer attributions.

**The quality of life of customers does not play a moderating role in the relationship between the alignment of the Bank's CSR activities with its business and attitudinal loyalty.**

Furthermore, we were unable to substantiate the moderating effect of quality of life on the relationship between the alignment of the Bank's CSR activities and its business and attitudinal loyalty. This situation contradicts our hypothesis, positing that quality of life moderates the relationship between the alignment of the Bank's CSR activities and its business and attitudinal loyalty. However, our structural model indicates that, despite lacking a moderating effect, quality of life can positively influence attitudinal loyalty in Bank XYZ customers. Lack of CSR fit as an antecedent to attitudinal loyalty contributes to this situation.

The research by Ogunmokun and Timur (2022) provides insightful observations regarding the relationship between Quality of Life (QoL) and the perception of a bank's Corporate Social Responsibility (CSR) activities. Specifically, individuals with higher QoL tend to view the alignment of a bank's CSR activities with its business through the lens of self-actualization and transcendence. This perspective is particularly relevant in our findings related to Bank XYZ, where a majority of the customers exhibit high QoL. As a result, these customers hold the bank to stricter standards in terms of CSR alignment, reflecting their advanced stages of self-fulfillment and broader societal concerns.

Further illuminating this dynamic, the study by Hani et al. (2021) identifies trust, respect, and reciprocity as the key antecedents of QoL. These dimensions offer a more comprehensive framework for understanding how individuals attain QoL. Trust, respect, and reciprocity not only foster personal well-being but also enhance the perception of a bank's CSR activities, influencing how customers perceive and interact with the bank's initiatives. Additionally, Hani et al. (2021) emphasize the importance of timing in measuring the impact of CSR alignment on consumer behavior. Whether their sustainability marketing messages are up to date or insensitive, by considering the appropriate timing for such evaluations, we can gain a clearer understanding of how CSR-fit influences consumer behavior. This aspect is crucial for accurately capturing the dynamics of CSR initiatives and their reception among customers with varying levels of QoL.

**There exists a positive relationship between the level of attitudinal loyalty and the likelihood of customers engaging in cross-buying behavior.**

Our research corroborates the hypothesis that a higher level of attitudinal loyalty increases the likelihood of cross-buying behavior among customers. This finding underscores the importance for Bank XYZ to focus on cultivating customer loyalty, as it directly influences the propensity for customers to engage in cross-buying. This relationship aligns with the insights provided by Reinartz et al. (2008), who posited that cross-buying is a consequential aspect of loyalty behavior. Supporting this notion, Bennett and Rundle-Thiele (2002) found that the concept of attitudinal loyalty explains a significant portion of the variance in purchasing behavior. This suggests that the mindset and emotional connection customers have with a bank significantly impact their buying decisions.

Moreover, this research contributes additional insights into the determinants of cross-buying behavior (CBB). While previous studies, such as the one by Soureli et al. (2008), have identified trust towards the bank and brand image as antecedents of CBB, our findings reveal that attitudinal loyalty is another crucial factor. This broadens the understanding of what drives cross-buying behavior, indicating that a customer's emotional and psychological commitment to a bank plays a vital role in their decision to purchase additional products or services.

**The level of attitudinal loyalty has a significant influence on the likelihood of customers participating in customer advocacy.**

Our research confirmed a significant link between attitudinal loyalty and customer advocacy at Bank XYZ. The findings support the hypothesis that higher levels of attitudinal loyalty led to increased customer advocacy. Loyal customers, emotionally and psychologically connected to the bank, are more likely to advocate for it actively. This advocacy manifests in various ways, such as word-of-mouth recommendations and social media endorsements. This relationship emphasizes the importance of cultivating strong customer relationships, as loyal customers often become brand ambassadors, enhancing the bank's reputation and reach. Thus, strategies focused on building customer loyalty are crucial, not just for immediate financial gains but for creating a loyal customer base that champions the bank's brand.

Our findings are supported by the research of Ogunmokun and Timur (2022), which similarly identifies the significant influence of attitudinal loyalty on the likelihood of customer advocacy in the same context. This corroboration underlines the crucial role of a customer's emotional and psychological commitment in fostering advocacy for a bank. Complementing this, the study by Bandyopadhyay and Martell (2007) also demonstrates that behavioral loyalty is influenced by attitudinal loyalty. Their research highlights the interconnection between these dimensions of loyalty, suggesting that deep-rooted emotional loyalty not only motivates customers to advocate for a bank but also influences their overall behavioral loyalty, encompassing their engagement with the bank's services and products.

**Cross-buying does not act as a mediating factor in the relationship between attitudinal loyalty and customer advocacy, indicating no influence of one on the other.**

Our research did not confirm the hypothesized mediating role of cross-buying between attitudinal loyalty and customer advocacy in the context of Bank XYZ. Contrary to expectations, the study showed that while both cross-buying and customer advocacy are linked to attitudinal loyalty, they do not follow a sequential pattern. Customers exhibited a direct transition to advocacy behaviors without necessarily going through a cross-buying phase. This observation challenges the traditional view of customer engagement as a linear process and suggests a more varied customer journey. Additionally, no significant direct relationship was found between cross-buying behavior and customer advocacy, indicating their independent operation despite both stemming from attitudinal loyalty. This finding calls for a reconsideration of how loyalty influences customer behaviors in banking.

Building upon the foundational work of Bhatnagar et al. (2019), who delineated that loyalty comprises both attitudinal and behavioral components, and our research further corroborates this perspective. Specifically, his study highlights that customer loyalty is positively related to Cross-Buying Behavior (CBB) and Customer Advocacy (CA). This finding aligns with our research, providing supportive evidence that emphasizes the integral role of attitudinal loyalty in influencing customer behaviors. Our study echoes this sentiment, demonstrating a clear connection between the level of attitudinal loyalty and the manifestation of behaviors such as CBB and CA, thereby reinforcing the multifaceted nature of customer loyalty within the banking sector (Bhatnagar et al., 2019).

**Financial literacy does not significantly affect the model.**

Our study utilized a multi-group modeling approach to assess the impact of financial literacy levels on Bank XYZ's customers. The analysis found no significant differences in the model's outcomes between groups with varying financial literacy levels. Minor variations in Quality-of-Life metrics were noted but lacked statistical significance. This finding challenges the assumption that financial literacy significantly influences banking customer behaviors and attitudes, indicating that other factors may be more influential in shaping these aspects at Bank XYZ.

Balasubramanian and Sargent's (2020) work on developing a more accurate financial behavior predictor, focusing on gaps, may offer valuable insights for our research. They suggest that such gaps could be more pronounced and thus more significant in models like ours, potentially revealing deeper insights into financial behaviors. This idea aligns with our findings, suggesting a need to explore these gaps further in future studies.

Complementing this, Grohmann's (2018) research in developing countries highlights a preference for savings accounts over more sophisticated financial products. This tendency is mirrored in Bank XYZ's operations in Indonesia, where simpler financial products are more prevalent. Additionally, Hsiao and Tsai (2018) observed that financial literacy can lower barriers, enabling people to purchase more sophisticated products. This could explain the variations in product choices we observed at Bank XYZ, suggesting that financial literacy plays a crucial role in influencing customer decisions about financial products.

In their study, Balasubramnian and Sargent (2020) developed a more significant predictor of financial behavior, focusing on the gaps between perceived and actual financial literacy. This approach suggests that examining the discrepancies between what customers believe they know and their actual knowledge might reveal more significant differences in our model. This concept aligns with the findings of Grohmann (2018), who noted that in developing countries, there is a tendency for customers to hold more savings accounts rather than engage with more sophisticated financial products. This pattern is also evident in the case of Bank XYZ, operating in Indonesia, where a preference for simpler financial products prevails among the customers. Furthermore, Hsiao and Tsai (2018) found that financial literacy lowers the barrier for people to purchase more sophisticated financial products. This evidence collectively supports our research findings, indicating that the role of financial literacy, and particularly the gap between perceived and actual literacy, may be a critical factor in understanding customer behavior towards different banking products in developing markets like Indonesia.

This study has limitations to consider. The limited number of respondents, due to time constraints, may affect the generalizability of the findings. A broader sample could offer deeper insights into customer behaviors and attitudes. Furthermore, using Structural Equation Modeling with Partial Least Squares (SEM-PLS) presents challenges. SEM-PLS is beneficial for complex models and non-normal data but lacks a global goodness-of-fit measure and may underestimate relationships. Its focus on prediction over theory testing and assumptions about measurement errors might influence the study's outcomes. These limitations should be considered in interpreting the study's results, impacting their robustness and applicability.

## VI. CONCLUSION AND IMPLICATIONS

Our research concludes that both cross-buying behavior and customer advocacy are direct outcomes of attitudinal loyalty at Bank XYZ. Key findings include: First, attitudinal loyalty significantly impacts cross-buying behavior and customer advocacy, underscoring the importance of emotional and psychological commitment; Second, customer advocacy at Bank XYZ does not solely stem from attitudinal loyalty via cross-buying, challenging traditional views of customer engagement; and third, financial literacy does not notably affect the relationship between CSR perception and customer behavior at Bank XYZ.

The study also explores the complexities of sustainability marketing in banking, emphasizing that effective CSR goes beyond mere implementation to how customers perceive these practices. This insight is crucial for enhancing both shareholder value and customer satisfaction. This research contributes to the field by providing a nuanced understanding of the loyalty-behavior nexus in banking, offering fresh perspectives on how loyalty influences customer actions independently of cross-buying. It suggests a need for reevaluation of traditional customer loyalty models in banking and calls for further exploration into the effects of attitudinal loyalty and the role of financial literacy. This study not only enriches academic discourse but also offers practical insights for the banking industry, aiming to improve customer satisfaction and business success.

**Appendix**

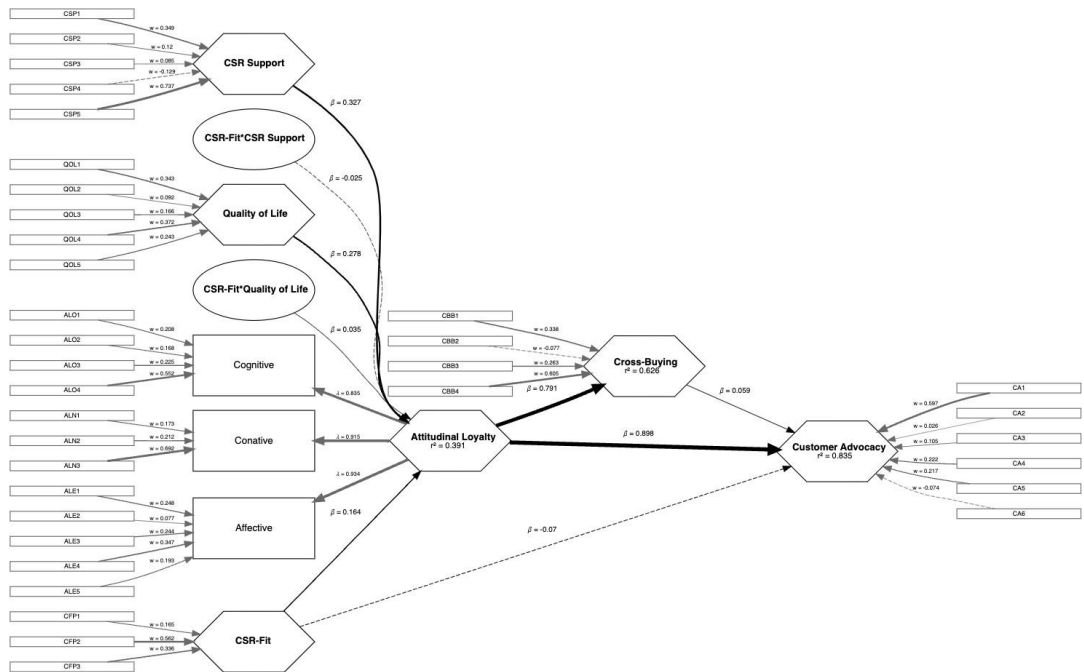

**Appendix 1: Structural Model**